\documentclass[
twocolumn,
preprintnumbers,amsmath,amssymb,prb
]{revtex4}

\usepackage{graphicx}
\usepackage{dcolumn}
\usepackage{bm}
\usepackage{amssymb}
\usepackage{epstopdf}
\usepackage{color}
\usepackage{amsmath}
\usepackage{ulem}

\begin{document}

\title{Optically induced delocalization of electrons bound by attractive potentials}
\author{O. V. Kibis}\email{Oleg.Kibis(c)nstu.ru}
\author{M. V. Boev}
\author{D. S. Eliseev}
\author{V. M. Kovalev}

\affiliation{Department of Applied and Theoretical Physics,
Novosibirsk State Technical University, Karl Marx Avenue 20,
Novosibirsk 630073, Russia}

\begin{abstract}

Within the Floquet theory of periodically driven quantum systems, we demonstrate that a circularly polarized off-resonant electromagnetic field can destroy the electron states bound by three-dimensional attractive potentials. As a consequence, the optically induced delocalization of bound electrons appears. The effect arises from the changing of topological structure of a potential landscape under a circularly polarized off-resonant electromagnetic field which turns simply connected potentials into doubly connected ones. Possible manifestations of the effect are discussed for conduction electrons in condensed-matter structures.

\end{abstract}

\maketitle

Controlling electronic properties of condensed-matter structures by a high-frequency off-resonant electromagnetic field, which is based on the Floquet theory of periodically driven quantum systems, has become an established research area~\cite{Oka_2019,Basov_2017,Goldman_2014,Bukov_2015,Eckardt_2015,Casas_2001,Kibis_2020_1,Nuske_2020,Kibis_2022,Liu_2022,Seshadri_2022,Kobayashi_2023}.
The off-resonant field cannot be absorbed by electrons and only dresses them, modifying electronic properties. Such a dressing results in many field-induced phenomena in various condensed-matter structures, including  semiconductor quantum wells~\cite{Lindner_2011,Kibis_2020_2,Kibis_2021_3}, quantum rings~\cite{Koshelev_2015}, quantum dots~\cite{Kryuchkyan_2017}, topological insulators~\cite{Rechtsman_2013,Wang_2013,Zhu_2023,Torres_2014}, carbon nanotubes~\cite{Kibis_2021_1},
graphene and related two-dimensional materials~\cite{Oka_2009,Syzranov_2013,Usaj_2014,Perez_2014,Sie_2015,Iurov_2019,Cavalleri_2020}, etc. Since all solids contain a lot of attractive potentials of various nature, there is a need to study electronic behavior in such a potential landscape under a high-frequency off-resonant electromagnetic field. In many previous studies on the subject, it was demonstrated both experimentally and theoretically that such a field shifts energy levels of electrons bound by attractive potentials due to the dynamical Stark effect (see, e.g., Refs.~\onlinecite{Delone_2000,Gavrila_1987}). However, the effect of the field on existence of the bound states still wait for detailed analysis. Solving this quantum-mechanical problem within the conventional Floquet theory, we found that a strong circularly polarized electromagnetic field can delocalize electrons bound by attractive potentials. The present Letter is dedicated to the first theoretical analysis of this all-optical mechanism of electron delocalization, which can manifest itself in various electronic systems.

\begin{figure}[!h]
\includegraphics[width=.8\columnwidth]{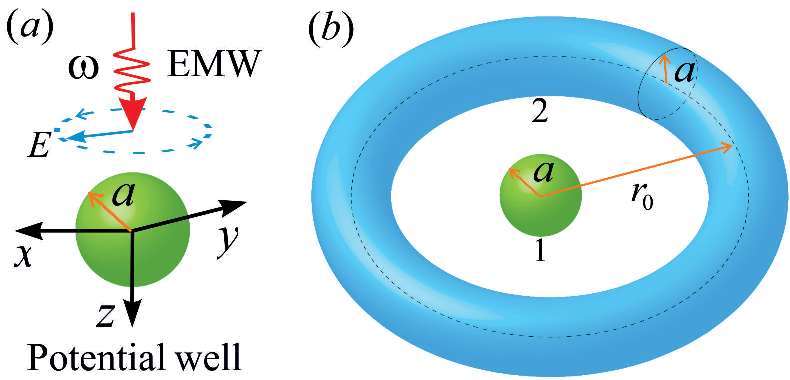}
\caption{Sketch of the system under consideration: (a) The potential well of radius $a$ irradiated by the circularly polarized electromagnetic wave (EMW) with the frequency $\omega$ and the electric field amplitude $E$; (b) The spherically symmetric potential well (1) transformed by the irradiation into the toroidal potential well (2), where $r_0$ is the radius of classical electron trajectory in the wave.}\label{Fig.1}
\end{figure}
Let us consider a potential well with the potential energy $U(\mathbf{R})$, where $\mathbf{R}=(x,y,z)$ is the radius vector, which is irradiated by a circularly polarized electromagnetic wave propagating along the $z$ axis (see Fig.~1a). Assuming that the wave length much exceeds the well size $a$, the interaction between an electron in the well and the wave can be described within the dipole approximation. Then the electron Hamiltonian reads
\begin{equation}\label{HA}
\hat{\cal H}_e=\frac{[\hat{\mathbf{p}}-e\mathbf{A}(t)/c]^2}{2m_e}+U(\mathbf{R}),
\end{equation}
where
\begin{equation}\label{A}
\mathbf{A}(t)=(A_x,A_y,A_z)=[cE/\omega](\sin\omega t,\,\cos\omega t,\,0)
\end{equation}
is the vector potential of the circularly polarized field, $\omega$ is the wave frequency assumed to be far from all resonant frequencies of the electron, $E$ is the electric field amplitude of the wave, $\hat{\mathbf{p}}=(\hat{p}_x,\hat{p}_y,\hat{p}_z)$ is the momentum operator, $m_e$ is the electron mass, and $e=-|e|$ is the electron charge. Let us apply the Kramers-Henneberger unitary
transformation,
\begin{equation}\label{KHU}
\hat{\cal
U}(t)=\exp\left\{\frac{i}{\hbar}\int^{\,t}\left[
\frac{e}{m_ec}\mathbf{A}(t^\prime)\hat{\mathbf{p}}-\frac{e^2}{2m_ec^2}A^2(t^\prime)
\right]dt^\prime\right\},
\end{equation}
which removes the coupling of the momentum $\hat{\mathbf{p}}$ to the vector potential $\mathbf{A}(t)$ in the Hamiltonian (\ref{HA}) and transfers the time dependence from the kinetic energy of electron to its potential energy~\cite{Kramers_52,Henneberger_68}. Then the transformed Hamiltonian (\ref{HA}) reads
\begin{eqnarray}\label{H}
\hat{\cal H}&=&\hat{\cal U}^\dagger(t)\hat{\cal
H}_e\hat{\cal U}(t) - i\hbar\hat{\cal U}^\dagger(t)\partial_t
\hat{\cal U}(t)\nonumber\\
&=&\frac{\hat{\mathbf{p}}^2}{2m_e}+U(\mathbf{R}-\mathbf{R}_0(t)),
\end{eqnarray}
where the radius vector $\mathbf{R}_0(t)=(r_0\cos\omega t,\,-r_0\sin\omega t,\,0)$ describes the classical circular trajectory
of electron movement under the circularly polarized field (\ref{A}), and
\begin{equation}\label{r0}
r_0=\frac{|e|E}{m_e\omega^2}
\end{equation}
is the radius of the trajectory~\cite{Landau_2}. Since the Hamiltonian \eqref{H} involves the only field-dependent parameter (\ref{r0}), this radius $r_0$ will be used in the problems analyzed below as a parameter describing the strength of electron-field interaction. Expanding the oscillating potential in the Hamiltonian (\ref{H}) into a
Fourier series, the Hamiltonian can be rewritten as
\begin{equation}\label{H3}
\hat{\cal
H}=\frac{\hat{\mathbf{p}}^2}{2m_e}+U_0(\mathbf{r})+\left[\sum_{n=1}^\infty
U_n(\mathbf{r})e^{in\omega t}+\mathrm{c.\,c.}\right],
\end{equation}
where
\begin{equation}\label{UR}
U_n(\mathbf{R})=\frac{1}{2\pi}\int_{-\pi}^{\pi}U\big(\mathbf{R}-\mathbf{R}_0(t)\big)e^{-in\omega t}\,d(\omega
t)
\end{equation}
are the harmonics of the Fourier expansion.
The Hamiltonian (\ref{H3}) is still physically equal to the initial Hamiltonian (\ref{HA}). Next, we need to make some approximations. Within the conventional Floquet theory of periodically driven quantum systems, one can introduce the unitary transformation $\hat{\cal U}_0(t)=e^{i\hat{S}(t)}$, which transforms the periodically time-dependent Hamiltonian (\ref{H3}) into the effective stationary Hamiltonian
\begin{equation}\label{H0}
\hat{\cal H}_{\mathrm{eff}}=\hat{\cal U}_0(t)^\dagger\hat{\cal H}\hat{\cal U}_0(t) -
i\hbar\hat{\cal U}_0^\dagger(t)\partial_t
\hat{\cal U}_0(t).
\end{equation}
There is the regular method to find the transformation operator $\hat{S}(t)$ in the case of high-frequency field. Namely, both the operator $\hat{S}(t)$ and the stationary Hamiltonian (\ref{H0}) can be found as an $1/\omega$-expansion (the Floquet-Magnus expansion)~\cite{Eckardt_2015,Goldman_2014,Bukov_2015,Casas_2001}, which leads to the effective stationary Hamiltonian
\begin{equation}\label{Hef}
\hat{\cal H}_{\mathrm{eff}}=\hat{\cal H}_{0}+\sum_{n=1}^\infty\frac{[\hat{\cal H}_{n},\hat{\cal H}_{-n}]}{n\hbar\omega}+{\it o}\left(\frac{1}{\omega}\right).
\end{equation}
In the high-frequency limit, one can restrict the expansion (\ref{Hef}) by its main term
\begin{equation}\label{H00}
\hat{\cal
H}_0=\frac{\hat{\mathbf{p}}^2}{2m_e}+U_0(\mathbf{R}),
\end{equation}
which will be under consideration in the following.

It should be noted that the effective stationary potential $U_0(\mathbf{R})$ in the Hamiltonian (\ref{H00}) has the clear physical meaning. In the labor reference frame, a free electron rotates along a circular trajectory with the radius (\ref{r0}) under the circularly polarized field (\ref{A}). The unitary transformation (\ref{KHU}) corresponds to transition from the labor reference frame to the rest frame of the rotating electron, where the potential well rotates along the circular trajectory with the field frequency. If the frequency is high enough, the electron ``feels'' only the  rotating potential $U(\mathbf{R}-\mathbf{R}_0(t))$ averaged over the rotation period $2\pi/\omega$, which is described by the stationary potential $U_0(\mathbf{R})$.

Let us consider a three-dimensional spherically symmetric attractive potential
\begin{equation}\label{UR}
U(\mathbf{R})=U(R)
\end{equation}
of the size $a$, which is significantly differs from zero only for $R<a$ (a short-range potential well). Then the effective potential reads
\begin{align}\label{U0}
&U_0(\mathbf{R})=\frac{1}{2\pi}\int_{-\pi}^{\pi}U\big(\mathbf{R}-\mathbf{R}_0(t)\big)\,d(\omega
t)\nonumber\\
&=\frac{1}{2\pi}\int_{-\pi}^{\pi}
U(\rho)d(\omega
t),
\end{align}
where
$$
\rho=\sqrt{(r-r_0)^2+z^2+4[(r-r_0)r_0+r_0^2]\sin^2(\omega t/2)},
$$
is the radius vector length in the coordinate system associated with the rotating potential, and $\mathbf{r}=(x,y)$ is the plane radius vector. In what follows, we will restrict the consideration by the case of large radius \eqref{r0} which meets the condition
\begin{equation}\label{C}
r_0\gg a.
\end{equation}
Since the rotating potential $U(\mathbf{R}-\mathbf{R}_0(t))$ significantly differs from zero only within the coordinate range $|r-r_0|<a$, the potential \eqref{U0} under the condition \eqref{C} can be rewritten as
\begin{equation}\label{U1}
U_0(r^\prime)=\frac{1}{2\pi}\int_{-\pi}^{\pi}
U\left(\sqrt{(r^\prime)^2+4r_0^2\sin^2(\omega t/2)}\right)\,d(\omega
t),
\end{equation}
where $r^\prime=\sqrt{(r-r_0)^2+z^2}$ is the radial coordinate of a torus with the radius $r_0$. Thus, the spherically symmetric potential well \eqref{UR} rotating along a circular trajectory of large radius \eqref{r0} turns into the effective toroidal potential well (\ref{U1}) pictured schematically in Fig.~1b.

Next, let us find electron states bound by the toroidal potential \eqref{U1}. The wave functions of the sought bound states can be written in the cylindrical coordinates $(z,r,\varphi)$ as $\Psi_m(z,r)e^{im\varphi}$ with $m=0,\pm1,\pm2,...$, where $\Psi_m(z,r)$ is the eigenfunction of the Schr\"odinger equation
\begin{align}\label{Sh10}
&\frac{\hbar^2}{2m_e}\left[\frac{\partial^2}{\partial r^2}+
\frac{1}{r}\frac{\partial}{\partial r}-\frac{m^2}{r^2}+\frac{\partial^2}{\partial z^2}+\varepsilon_m\right]\Psi_0(z,r)\nonumber\\
&=U_0(r^\prime)\Psi_m(z,r),
\end{align}
and $\varepsilon_m$ is the energy of the sought bound state. At the current stage of consideration, let us omit the second term in the square brackets of Eq.~(\ref{Sh10}). Physically, such an approximation corresponds to neglecting curvature of the toroidal potential well. The approximation is correct if the torus radius $r_0$ is large enough, what will be justified below.
Under this approximation, the three-dimensional Schr\"odinger equation \eqref{Sh10} for the ground bound state $(m=0)$ reduces to the two-dimensional equation,
\begin{equation}\label{Sh3}
-\frac{\hbar^2}{2m_e}\left[\frac{\partial^2}{\partial {x^\prime}^2}+\frac{\partial^2}{\partial {y^\prime}^2}\right]\Psi_0(\mathbf{r}^\prime)+U_0(r^\prime)\Psi_0(\mathbf{r}^\prime)
=\varepsilon_0\Psi_0(\mathbf{r}^\prime),
\end{equation}
where $x^\prime=r-r_0$ and $y^\prime=z$ are the new coordinates, and $\mathbf{r}^\prime=(x^\prime,y^\prime)$ is the radius vector written in these coordinates. Next, let us introduce the polar coordinates $(r^\prime,\theta)$, where the radial coordinate is $r^\prime=\sqrt{{x^\prime}^2+{y^\prime}^2}=\sqrt{(r-r_0)^2+z^2}$ and the azimuthal coordinate is $\theta(z,r)=\arctan({x^\prime}/{y^\prime})=\arctan([r-r_0]/z)$. Then eigenfunctions of the Schr\"odinger problem \eqref{Sh3} can be written as $\Psi_0(\mathbf{r}^\prime)=\psi_{m^\prime}(r^\prime)e^{im^\prime\theta(z,r)}$ with $m^\prime=0,\pm1,\pm2,...$, where the wave function corresponding to the sought ground bound state $(m^\prime=0)$ satisfies the equation
\begin{equation}\label{Sh2}
-\frac{\hbar^2}{2m_e}\left[\frac{\partial^2}{\partial {r^\prime}^2}+
\frac{1}{r^\prime}\frac{\partial}{\partial r^\prime}\right]\psi_0(r^\prime)+U_0(r^\prime)\psi_0(r^\prime)
=\varepsilon_0\psi_0(r^\prime).
\end{equation}
Since the depth of the toroidal potential well \eqref{U1} decreases with increasing the torus radius $r_0$, it is shallow under the condition \eqref{C}. The solution of Eq.~\eqref{Sh2} for such a shallow two-dimensional well is well-known. Following Landau and Lifshitz~\cite{Landau_3}, Eq.~\eqref{Sh2} yields the ground bound state with the binding energy
\begin{equation}\label{e}
|\varepsilon_0|\sim\frac{\hbar^2}{m_ea^2}\exp\left[-\frac{\hbar^2}{m_e}
\left|\int_0^\infty U_0(r^\prime)r^\prime\,dr^\prime\right|^{-1}\right].
\end{equation}
and the wave function $\psi_0(r^\prime)$ which is approximately equal to a constant inside the potential well $U_0(r^\prime)$ and decreases outside the well as the Hankel function $H_0(i\varkappa_0r^\prime)$, where $\varkappa_0=\sqrt{2m_e|\varepsilon_0|/\hbar^2}\gg1/a$ is the inverse localization scale of the bound state. Substituting the found wave function $\psi_0(r^\prime)$ into Eq.~(\ref{Sh10}), one can see that the omitted second term in the square brackets contributes with the smallness $\sim1/\varkappa_0r_0$ if $\varkappa r_0\gg1$. Thus, Eq.~\eqref{e} correctly describes the bound state under the condition
\begin{equation}\label{c3}
\varkappa_0r_0\gg 1.
\end{equation}
The exponential decreasing of the binding energy \eqref{e} with increasing the radius \eqref{r0} suggests that the bound states of the toroidal well \eqref{U1} disappear at some critical value of the radius $r_0$ beyond applicability of the condition \eqref{c3}. To prove this guess, the exact Schr\"odinger equation \eqref{Sh10} should be solved accurately as follows.

Since the toroidal potential well \eqref{U1} is shallow under the condition \eqref{C}, it can contain only bound states whose localization scale much exceeds the potential scale, $\varkappa_0a\ll1$. Then one can make the following replacement in the right side of Eq.~\eqref{Sh10}, $$U_0(r^\prime)\Psi_m(z,r)\rightarrow U_0(r^\prime)\Psi_m(0,r_0).$$
Applying the Fourier transformation to the wave functions of the bound states,
\begin{equation}\label{Sh3.0}
\Psi_m(z,r)=\int\limits_{-\infty}^{\infty}\frac{dq}{2\pi}e^{iqz}\psi_m(q,r),
\end{equation}
we arrive from Eq.~(\ref{Sh10}) at the Schr\"odinger equation in the $q$-representation,
\begin{align}\label{Sh3.1}
&\left[\frac{\partial^2}{\partial r^2}+\frac{1}{r}\frac{\partial}{\partial r}-
\left(\varkappa_m^2+q^2+\frac{m^2}{r^2}\right)\right]\psi_m(q,r)\nonumber\\
&=\frac{2m_e}{\hbar^2}\,u(q,r)\Psi_m(0,r_0),
\end{align}
where $\varkappa_m=\sqrt{2m_e|\varepsilon_m|/\hbar^2}$ is the inverse localization scale of the bound state with the energy $\varepsilon_m$, and $u(q,r)=\int_{-\infty}^\infty dz\,e^{-iqz}U_0(r^\prime)$ is the Fourier image of the potential \eqref{U1} along the $z$ axis. The localized eigenfunctions of Eq.~(\ref{Sh3.1}), which turn into zero at $r\rightarrow\infty$, can be written as
\begin{equation}\label{Sh3.2}
\psi_m(q,r)=\left\{
            \begin{array}{ll}
              A(q)I_m\left(r\sqrt{\varkappa_m^2+q^2}\right), & {r_0-r\ll a} \\
              B(q)K_m\left(r\sqrt{\varkappa_m^2+q^2}\right), & {r-r_0\gg a}
            \end{array}
          \right.,
\end{equation}
where $I_m(x)$ and $K_m(x)$ are the modified Bessel functions of the first and second kind (the Infeld and MacDonald functions, respectively), whereas $A(q)$ and $B(q)$ are some coefficients. To join the two solutions \eqref{Sh3.2}, one can apply the known approach to solve the Schr\"odinger problem with a shallow potential well~\cite{Landau_3}. Namely, let us introduce the two points, $r=r_0\pm\bar{r}$, which satisfy the condition $a\ll\bar{r}\ll1/\varkappa_m$. Then the continuity conditions for the wave function (\ref{Sh3.2}) at these two points yield the equalities
\begin{eqnarray}\label{cc1}
A(q)I_m\left(r_0\sqrt{\varkappa_m^2+q^2}\right)&=&\Psi_m(0,r_0),\nonumber\\ B(q)K_m\left(r_0\sqrt{\varkappa_m^2+q^2}\right)&=&\Psi_m(0,r_0).
\end{eqnarray}
Integrating Eq.~(\ref{Sh3.1}) between these two points over $r$, one can obtain another equality,
\begin{align}\label{cc2}
&B(q)K_m^\prime\left(r_0\sqrt{\varkappa_m^2+q^2}\right)-A(q)I_m^\prime\left(r_0\sqrt{\varkappa_m^2+q^2}\right)\nonumber\\
&=\frac{2m_e}{\hbar^2}\frac{\Psi_m(0,r_0)}{\sqrt{\varkappa_m^2+q^2}}\int_{r_0-\bar{r}}^{r_0+\bar{r}}u(q,r)dr\nonumber\\
&\approx\frac{2m_e}{\hbar^2}\frac{\Psi_m(0,r_0)}{\sqrt{\varkappa_m^2+q^2}}\int_{0}^{\infty}u(q,r)dr,
\end{align}
where $I'_{m}(x)\equiv dI_{m}(x)/dx$, $K'_{m}(x)\equiv dK_{m}(x)/dx$. As a result, we arrive at the algebraic system of the two equations, which yields
\begin{align}\label{App3}
&\left[
  \begin{array}{c}
    A(q) \\
    B(q) \\
  \end{array}
\right]=-\left[
  \begin{array}{c}
    K_{m}\left(r_0\sqrt{\varkappa_m^2+q^2}\right) \\
    I_{m}\left(r_0\sqrt{\varkappa_m^2+q^2}\right) \\
  \end{array}
\right]\nonumber\\
&\times\frac{\Psi(0,r_0)}{{D}(q)\sqrt{\varkappa_m^2+q^2}}
\frac{2m_e{u}_0(q)}{\hbar^2},
\end{align}
where
\begin{eqnarray}\label{D}
{D}(q)&=&I'_{m}\left(r_0\sqrt{\varkappa_m^2+q^2}\right)K_{m}\left(r_0\sqrt{\varkappa_m^2+q^2}\right)\nonumber\\
&-&K'_{m}\left(r_0\sqrt{\varkappa_m^2+q^2}\right)I_{m}\left(r_0\sqrt{\varkappa_m^2+q^2}\right)
\end{eqnarray}
is the determinant of the system, and
\begin{equation}\label{uq}
u_0(q)=\int_{-\infty}^\infty{dz}\int_0^\infty dr\,e^{-iqz}U_0(r^\prime)
\end{equation}
is the Fourier image of the potential \eqref{U1} along the $z$ axis averaged in the $(x,y)$ plane.
Applying the known relations for the modified Bessel functions, $2K'_m(x)=-[K_{m+1}(x)+K_{m-1}(x)]$,
$2I'_m(x)=I_{m+1}(x)+I_{m-1}(x)$, and $I_m(x)K_{m+1}(x)+I_{m+1}(x)K_m(x)=1/x$, the determinant (\ref{D}) reads ${D}(q)=[r_0\sqrt{\varkappa_m^2+q^2}]^{-1}$. Then Eqs.~\eqref{Sh3.0}--\eqref{uq} yield the wave function
\begin{align}\label{App4}
&\Psi_m(z,r)=-\Psi_m(0,r_0)\frac{2m_er_0}{\hbar^2}
\int_{-\infty}^{\infty}\frac{dq}{2\pi}e^{iqz}u_0(q)\nonumber\\
&\times
\left\{
            \begin{array}{ll}
              I_m\left(r\sqrt{\varkappa_m^2+q^2}\right)K_{m}\left(r_0\sqrt{\varkappa_m^2+q^2}\right), & r\leq r_0 \\
              K_m\left(r\sqrt{\varkappa_m^2+q^2}\right)I_{m}\left(r_0\sqrt{\varkappa_m^2+q^2}\right), & r\geq r_0
            \end{array}
          \right.,
\end{align}
where the constant $\Psi_m(0,r_0)$ can be found from the normalization condition, $2\pi\int_0^\infty rdr\int_{-\infty}^\infty dz\,|\Psi_m(z,r)|^2=1$. Substituting $z=0$ and $r=r_0$ into Eq.~\eqref{App4}, we arrive at the integral equation defining the energy spectrum of the bound states,
\begin{align}\label{Sh3.3}
&\int_{-\infty}^{\infty}\frac{dq}{2\pi}K_m\left(r_0\sqrt{\varkappa_m^2+q^2}\right) I_m\left(r_0\sqrt{\varkappa_m^2+q^2}\right){u}_0(q)\nonumber\\
&=-\frac{\hbar^2}{2m_er_0},
\end{align}
where the index $m=0$ corresponds to the ground bound state.

To solve Eq.~\eqref{Sh3.3} analytically, there is a need to make some approximations. First, it should be noted that the Fourier image \eqref{uq} for any short-range potential of the size $a$ can be written approximately as
\begin{equation}\label{uq000}
u_0(q)\approx\left\{
            \begin{array}{ll}
              u_0(0), & {|q|\leq1/a} \\
              0, & {|q|>1/a}
            \end{array}
          \right.,
\end{equation}
where
\begin{align}\label{uq0}
&{u}_0(0)=\int_{-\infty}^\infty{dz}\int_0^\infty dr\,U_0(r^\prime)=\int_{-\infty}^\infty{dy^\prime}\int_{-r_0}^\infty dx^\prime\,U_0(r^\prime)\nonumber\\
&\approx{2\pi}\int_{0}^\infty U_0(r^\prime)r^\prime{dr^\prime},
\end{align}
Second, let us assume the condition \eqref{c3} to be satisfied. Then, using the known asymptotic expressions for the modified Bessel functions at their large arguments and evaluating the integral in Eq.~\eqref{Sh3.3} with the logarithmic accuracy, we arrive from Eqs.~\eqref{Sh3.3}--\eqref{uq0} with $m=0$ at the transcendental equation,
\begin{equation}\label{ln}
\ln\left[\frac{1+\sqrt{1+(\varkappa_0a)^2}}{\varkappa_0a}\right]=\frac{\hbar^2}{2m_e}\left|\int_0^\infty U_0(r^\prime)r^\prime\,dr^\prime\right|^{-1}.
\end{equation}
This equation yields the binding energy of the ground bound state, $\varepsilon_0$, which, as expected, exactly coincides with the energy \eqref{e} derived above from the approximate Schr\"odinger equations \eqref{Sh3}--\eqref{Sh2} under the same condition \eqref{c3}.

The Floquet function, which is the eigenfunction of the periodically time-dependent Hamiltonian (\ref{HA}) and describes the found bound state \eqref{e} in the labor reference frame, reads
\begin{equation}\label{F}
F_0(\mathbf{R},t)=e^{-i\varepsilon_0 t/{\hbar}}\,\hat{\cal U}(t)\Psi_0(z,r).
\end{equation}
It should be noted that the term $(e^2/2m_ec^2)A^2(t^\prime)$ in the unitary transformation (\ref{KHU}) leads only to the energy shift of all electron states by the energy of electron rotation under the field, $E^2/2m_e\omega^2$. Therefore, it does not affect electronic properties and can be omitted. As a result, the unitary transformation $\hat{\cal U}(t)$ in Eq.~(\ref{F}) yields only the coordinate replacements $x\rightarrow x+r_0\cos\omega t$ and $y\rightarrow y-r_0\sin\omega t$ in the wave function \eqref{App4} with $m=0$.

It follows from Eqs.~\eqref{Sh3.3}--\eqref{uq0} that bound states in the toroidal well \eqref{U1} disappear for $r_0\geq \rho_0$, where the critical radius $r_0=\rho_0$ corresponds to the zero binding energy of the ground bound state ($\varkappa_0=0$) and is defined by the integral equation
\begin{equation}\label{Sh3.4}
\int_{0}^{\rho_0/a}K_0(x)I_0(x)\,{dx}
=\frac{\hbar^2}{4m_e}\left|\int_0^\infty U_0(r^\prime)r^\prime\,dr^\prime\right|^{-1}_{r_0=\rho_0}.
\end{equation}
As a consequence, the field-induced delocalization of electrons bound by the potential \eqref{UR} appears if the field \eqref{A} is strong enough to satisfy the condition $r_0\geq\rho_0$.

The theory presented above was developed for the potential well \eqref{UR} of most general form. To proceed, one needs to  apply this theory to some model potential. For definiteness, let us consider the Gaussian potential well,
\begin{equation}\label{GP}
U(R)=-|V|\exp(-{R^2}/a^2),
\end{equation}
which always contains bound electron states under the condition $|V|>\hbar^2/m_ea^2$ assumed to be satisfied. Substituting the potential \eqref{GP} into Eq.~\eqref{U0} and evaluating integral there with the saddle-point method, we arrive at the effective potential,
\begin{equation}\label{U11}
U_0(r^\prime)=-\frac{|V|a}{2\sqrt{\pi} r_0}\exp(-{r^\prime}^2/a^2),
\end{equation}
which contains the ground bound state with the binding energy \eqref{e},
\begin{equation}\label{e0}
|\varepsilon_0|\sim\frac{\hbar^2}{m_ea^2}\exp\left[-\frac{4\sqrt{\pi}\hbar^2r_0}{m_e|V|a^3}\right],
\end{equation}
under the condition \eqref{c3}. One can see that both the depth of the toroidal potential well \eqref{U11} and the binding energy \eqref{e0} decrease with increasing the ratio $r_0/a$ according to the general theory developed above. To complete the analysis, the integral equation \eqref{Sh3.3} with the toroidal potential \eqref{U11} was solved numerically. It follows from the solving that the binding energy of the ground bound state decreases with increasing $r_0$ (see Fig.~2a) and turns into zero at the critical radius $r_0=\rho_0$, which is plotted in Fig.~2b as a function of the well depth $|V|$.
\begin{figure}[!h]
\includegraphics[width=.7\columnwidth]{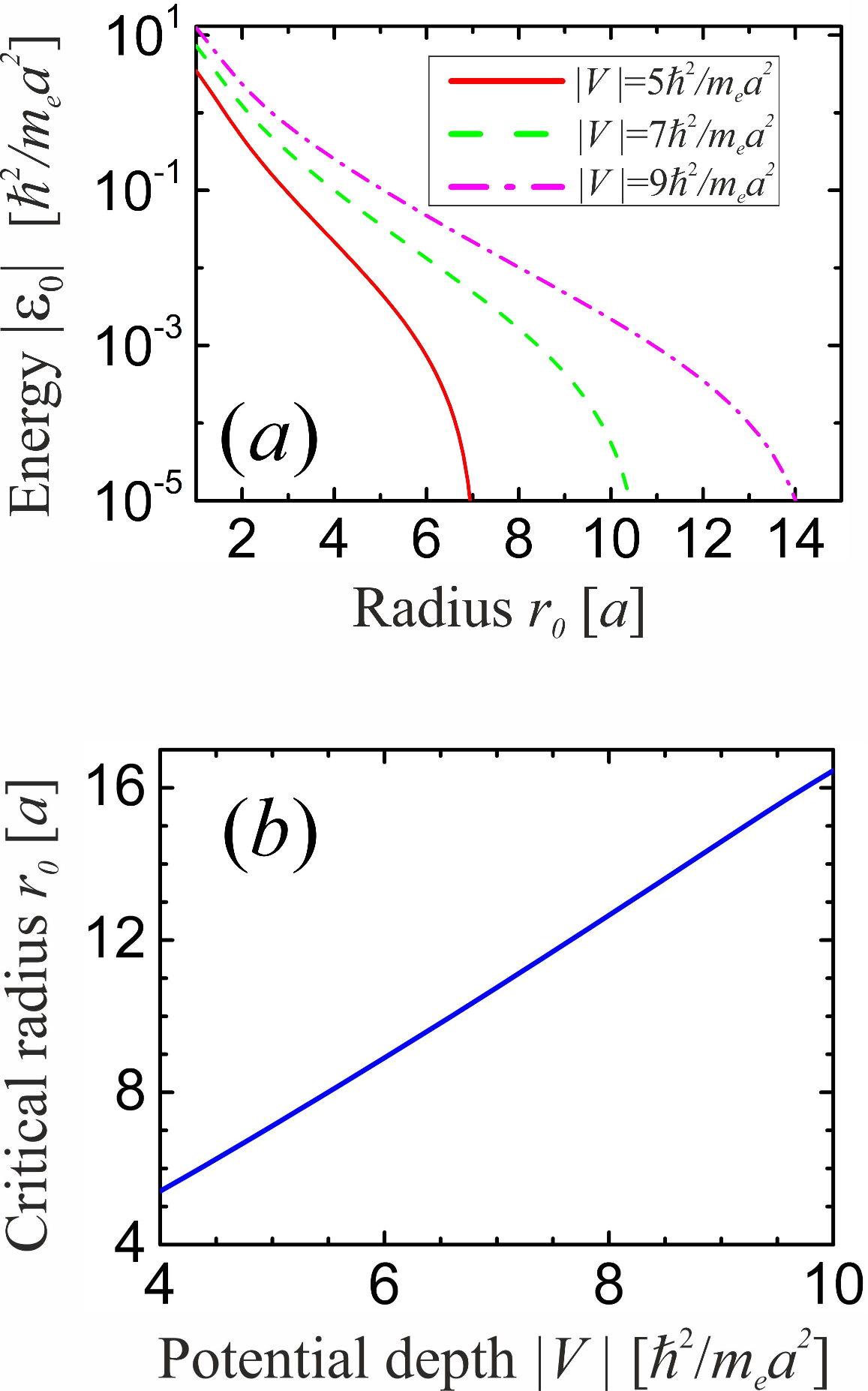}
\caption{Structure of bound states in the Gaussian potential well of the size $a$: (a) Dependence of the binding energy of the ground bound state, $|\varepsilon_0|$, on the radius $r_0$ for the different well depths $|V|$; (b) Dependence of the critical radius $r_0$ on the well depth $|V|$.}\label{Fig.2}
\end{figure}

It should be noted that the spherically symmetric attractive potential \eqref{UR} is the simple model to demonstrate the discussed effect with the pen-and-paper calculations. Going in the same way, there is no problem to analyze the effect for more realistic potentials with numerical simulations. However, this simple model is applicable, particularly, to describe accurately the physically important case of electrons bound by donor impurities in semiconductor materials with isotropic conduction band (e.g., GaAs). Since the localization radius of bound electrons much exceeds the crystal lattice spacing in such materials, the Schr\"odinger equation for a bound electron can be written in the conventional effective mass approximation, where the attractive potential of a donor is spherically symmetric (the screened Coulomb potential). To observe the discussed effect in the bulk of a material, the screening length of a high-frequency field should be large enough for the material. Therefore, semiconductor materials with the low density of conduction electrons (i.e., with the large screening length) are preferable from experimental viewpoint.

The small parameter of the series expansion (9) is the ratio of the binding energy of an electron bound at an attractive potential, $|\varepsilon_0|$, and the photon energy, $\hbar\omega$. Thus, the developed theory is correct if this ratio satisfies the condition $|\varepsilon_0|/\hbar\omega\ll1$. Since electrons bound by shallow impurities in semiconductors have the binding energy of meV scale, the high-frequency fields around (and above) the THz frequency range can be used to induce the considered effect. Using the size $a\sim10$~nm, which is typical for a shallow potential landscape in semiconductor materials, the critical value of the radius \eqref{r0} can be estimated as tens of nm. Then it follows from Eq.~\eqref{r0}, particularly, that the field with the frequency $\sim10$~THz and the electric field amplitude $E\sim10^{2}$~V/$\mu$m --- which is achievable in state-of-the-art experiments on the Floquet engineering of condensed-matter structures (see, e.g., Ref.~\onlinecite{Cavalleri_2020}) --- is appropriate to observe the discussed effect.

Accomplishing the discussion, it should be noted that a circularly polarized electromagnetic field changes the topological structure of quantum well, transforming the simply connected spherical well \eqref{UR} into the doubly connected toroidal well \eqref{U1}. Such a topological phase transition is accompanied by the crucial modification of electronic properties. As a main result, the doubly connected toroidal well \eqref{U1} loses the bound states which take place in the simply connected spherical well \eqref{UR}. It should be noted that a linearly polarized high-frequency field --- in contrast to circularly polarized one --- does not change the topological structure of potentials. As a consequence, the approach developed above is not applicable to describe the delocalization effect induced by a linearly polarized field. Moreover, it is known that electron states bound by an attractive Coulomb potential (hydrogen atom) irradiated by a linearly polarized field remain localized for any field amplitude and frequency~\cite{Gavrila_1987}. Therefore, the question about possibility of the delocalization of bound electrons under a linearly polarized field cannot be answered in general form. This problem is still opened for discussion and needs numerical simulations for a specific potential to be solved properly.

Concluding, it follows from the present analysis that various attractive potentials --- which are normally contain bound electron states --- can lose them under irradiation by a circularly polarized off-resonant electromagnetic field. As a consequence, the optically induced delocalization of bound electrons appears. This effect arises from changing topological structure of a potential landscape under the field and can manifest itself in various electronic systems. Among them, conducting condensed-matter structures should be noted especially. Normally, they contain a lot of attractive potentials which capture electrons. It follows from the present theory that a circularly polarized field can delocalize captured electrons, resulting in increasing density of conduction electrons. As a consequence, one can expect the experimentally observable increasing conductivity under the field.

{\it Acknowledgments.}
The reported study was funded by the Russian Science Foundation (project 20-12-00001).

\end{document}